\documentclass[prb,aps,twocolumn]{revtex4}
\usepackage{bm}
\usepackage{epsfig}
\usepackage{latexsym}
\usepackage{amssymb}
\usepackage{amsmath}
\usepackage{graphicx}
\usepackage{color}

\begin{document}

\title{Enhanced spin relaxation time due to electron-electron scattering in
semiconductors}
\author{W.J.H.~Leyland}\thanks{Permanent address: Cavendish Laboratory, Madingley Road, Cambridge CB4
3HE, UK.}
\author{G.H.~John and R.T.~Harley} \affiliation{School of
Physics and Astronomy, University of Southampton, Southampton SO17
1BJ, UK.}
\author{M.M.~Glazov and E.L.~Ivchenko}
\affiliation{A.F.Ioffe Physico-Technical Institute, Russian
Academy of Sciences, 194021 St Petersburg, Russia.}
\author{D.A.~Ritchie}
\affiliation{Cavendish Laboratory, Madingley Road, Cambridge CB4
3HE,UK.}
\author{A.J.~Shields}
\affiliation{Toshiba Research Europe Ltd, Milton Road Science
Park, Cambridge CB4 4WE, UK}
\author{M.~Henini}
\affiliation{School of Physics and Astronomy, University of
Nottingham, Nottingham NG7 4RD, UK}

\begin{abstract}
We present a detailed experimental and theoretical analysis of the
spin dynamics of two-dimensional electron gases (2DEGs) in a
series of $n$-doped GaAs/AlGaAs quantum wells.
Picosecond-resolution polarized pump-probe reflection techniques
were applied in order to study in detail the temperature-,
concentration- and quantum-well-width- dependencies of the spin
relaxation rate of a small photoexcited electron population. A
rapid enhancement of the spin life-time with temperature up to a
maximum near the Fermi temperature of the 2DEG was demonstrated
experimentally. These observations are consistent with the
D'yakonov-Perel' spin relaxation mechanism controlled by
electron-electron collisions. The experimental results and
theoretical predictions for the spin relaxation times are in good
quantitative agreement.
\end{abstract}

\date{\today}
\maketitle

\section{Introduction}\label{sec:1}

Expectations for device applications of non-equilibrium spin
populations of electrons and/or nuclei in semiconductors will
become more realistic when there is complete understanding of the
microscopic mechanisms which control spin coherence and
relaxation. This applies particularly to quantum well structures
as they are likely to be an important part of any such spintronic
device. In this paper we report an experimental and theoretical
study of electron spin relaxation at temperatures between 5 K and
300 K in a series of GaAs/AlGaAs quantum wells containing high
mobility two-dimensional electron gases (2DEGs). We have
previously published detailed accounts of our theoretical
approach~\cite{1,2,3} and preliminary accounts of the experiments
on one of the samples~\cite{4}. By fully characterizing the
electron mobility and concentration and making use of results from
an experimental study of low temperature spin dynamics in the same
samples, to be published separately~\cite{5}, we are able here to
give a complete quantitative theoretical description of the spin
dynamics in all the samples.

Three mechanisms are known for spin relaxation of electrons in
zinc-blende semiconductors~\cite{6,7}. Two usually make minor
contributions; these are spin-flips associated with electron
scattering due to spin-orbit interaction, the Elliott-Yafet (EY)
mechanism~\cite{8,9} and spin-flips induced by exchange
interaction with holes, the Bir-Aronov-Pikus (BAP)
mechanism~\cite{10}. The third, the D'yakonov-Perel' (DP)
mechanism~\cite{11,12} is the most important particularly in
$n$-type samples. In the DP mechanism the driving force for spin
reorientation is the intrinsic tendency of electron spins to
precess in the effective magnetic field which they experience as a
result of spin-orbit interaction. This is quantified as the
spin-splitting of the conduction band. The corresponding
precession vector $\bm \Omega_{\bm k} $ varies in magnitude and
direction according to the electron wavevector $\bm k$. Under
normal conditions in a 2DEG a collision-dominated regime holds in
which the electron spin precession is frequently interrupted by
scattering causing spin reorientation to proceed as a succession
of randomly directed small fractional rotations. Approach to
equilibrium is exponential with spin relaxation rate along a
particular (main) axis of the structure, $i$, given by \cite{11,7}
\begin{equation}\label{eq1}
\tau _{s,i}^{-1} =\left\langle {\bm \Omega_{\bot}^2 }
\right\rangle \tau _p^\ast  \quad \quad \quad \quad (\left\langle
{\left|  {\bm \Omega_{\bot} } \right|} \right\rangle \tau _p^\ast
\ll 1),
\end{equation}
where $\left\langle {\bm \Omega _\bot ^2 } \right\rangle $ is the
square of the component of  $\bm \Omega_{\bm k} $ perpendicular to
the axis $i$ averaged over the spin-oriented population and $\tau
_p^\ast $ is the momentum scattering time of a \textit{single}
electron \cite{1}. Equation (\ref{eq1}) reflects the diffusive
character of the spin decoherence; the spin pseudovector performs
a random walk on the surface of a sphere and its displacement from
the initial position during a time $t$ is proportional to $\left[
{\left\langle {\bm \Omega _\bot ^2 } \right\rangle \tau _p^\ast t}
\right]^{\frac{1}{2}}$ ($\tau _p^\ast \ll t\ll \tau _s )$.

Equation (\ref{eq1}) contains the `motional slowing'
characteristic of the DP mechanism; scattering actually inhibits
spin reorientation so that increasing electron scattering produces
slower spin relaxation. In the past it was assumed~\cite{11,12,7}
that $\tau _p^\ast $ could be equated to the momentum relaxation
time of the electron ensemble, $\tau_{p}$, obtained from the
electron mobility. In previous papers, Refs.~\onlinecite{1,2,3,4},
we have pointed out that this assumption is invalid in
high-mobility $n$-type semiconductors. Furthermore, in 2DEGs, at
low temperatures we have shown \cite{4} that the
collision-dominated regime breaks down to give oscillatory
\cite{13} rather than exponential spin evolution. Here we
concentrate on spin dynamics of 2DEGs where the spin evolution is
exponential and therefore the assumption of strong scattering is
valid. We directly observe motional slowing and demonstrate that
$\tau _p^\ast $ is, in general, much shorter than $\tau_{p}$. Our
theoretical analysis shows that this is a result of
electron-electron scattering which can randomize spin precession
whilst having  almost no effect on the mobility. The conclusion is
that, except at very low temperatures, electron-electron
scattering is dominant in determining the spin dynamics in these
2DEGs. Furthermore it \textit{increases} the relaxation time above
that expected on the basis of scattering processes which limit the
electron mobility.

\section{Experimental techniques and results}\label{sec:2}

The samples (see Table 1) used for our experiments were two series
of one-side $n$-modulation doped single GaAs/AlGaAs quantum wells
of widths $L_{z}$ ranging from 6.8 nm to 20 nm and grown by MBE on
(001)-oriented semi-insulating GaAs substrates. The first series
designated T (for Toshiba) comprising nominal well widths of 20 nm
and 10 nm consisted of the following layers: substrate, 1 micron
GaAs, 1 micron AlGaAs, superlattice of 100 repeats of 2.5 nm GaAs
and 2.5 nm AlGaAs, GaAs quantum well, 60 nm AlGaAs, 200 nm
10$^{17}$ cm$^{-3}$ Si-doped AlGaAs and 17 nm GaAs cap layer. The
second series designated NU (Nottingham) with nominal well widths
10.2 nm and 6.8 nm had a different structure giving somewhat
higher electron concentrations: substrate, 2 microns GaAs, 10.2 nm
AlGaAs, seven repeats of 3.4 nm GaAs and 10.2 nm AlGaAs, GaAs
quantum well, 30.4 nm AlGaAs, 30.4 nm 10$^{18}$ cm$^{-3}$ Si-doped
AlGaAs and 25.4 nm GaAs cap layer. Except where specified all the
layers were undoped and the Al fractional concentration was 0.33.
They were fabricated into FET devices with transparent Schottky
gate for optical measurements and Hall contacts to the conducting
channel to allow control and \textit{in situ} measurements of Hall
mobility, $\mu $, and electron concentration, $N_{S}$. The bias
was set for maximum $N_{S}$ which also corresponded to maximum
$\mu $ in the wells. For low temperature measurements a liquid
helium flow cryostat was used in which cold gas surrounded the
sample.

Figure 1 shows Hall mobility measurements under the conditions of
bias and illumination used for the optical investigation of
spin-dynamics plotted as the corresponding ensemble momentum
scattering time \textit{$\tau $}$_{p }=$ \textit{$\mu $e/m}$_{e}$
where $e$ is the elementary charge and $m_{e}$ is the electron
effective mass. The values at the lowest temperatures are typical
for high quality single quantum wells, as opposed to
heterojunctions, with ensemble momentum relaxation times in the
range 10 ps to 27 ps (see Table 1). These values are probably
limited by neutral impurity and interface scattering. At higher
temperatures the mobility falls off in a manner consistent with
the onset of phonon scattering processes. The electron
concentrations obtained from Hall measurements for each sample are
constant up to at least 100 K and then fall off at higher
temperatures. Above 100 K, however, the measurements become
increasingly unreliable due to the possible existence of parallel
conduction paths within the sample associated with thermally
excited carriers. In our analysis (section \ref{sec:3}) we have
used the measured values of mobility and assumed $N_{S}$ to be
temperature independent; as described below, we have used optical
spectroscopy to determine the absolute value of $N_{S}$ and hence
Fermi temperature $T_{F}=E_{F}/k_{B}$, where $E_{F}$ is the Fermi
energy and $k_{B}$ is the Boltzmann constant (Table 1).

\begin{figure}[htbp]
\includegraphics[width=0.8\linewidth]{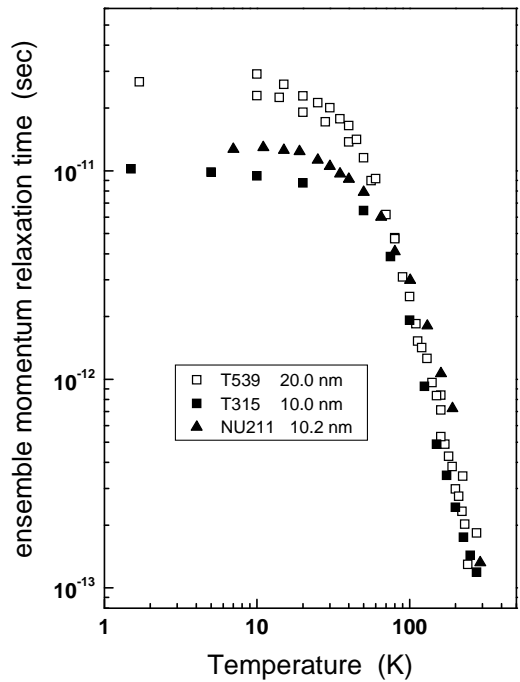}
\caption{Temperature dependence of the electron momentum
relaxation time (from mobility) for different samples. The
mobility in NU535 sample is essentially the same as in NU211
sample and is not shown here.} \label{fig1}
\end{figure}

\begin{table*}[htbp]
\caption{Sample paramters}
\begin{center}
\begin{tabular}{|c|p{59pt}|p{59pt}|p{58pt}|p{64pt}|p{61pt}|p{66pt}|p{59pt}|}
\hline & Nominal well width $L_{z}$ \par (nm)& Electron
confinement energy, $E_{e1}$ (meV)& Electron density \par $N_{S}$
\par (cm$^{-3})$& Fermi temperature \par $T_{F}$ \par (K)&
Ensemble momentum relaxation time $\tau_{p }$~(ps) at 5K& $\Omega
$(k$_{F})$ [5]
\par (rad ps$^{-1})$&
Single electron momentum relaxation time $\tau _p^\ast$~(ps) \par at 5K [5] \\
\hline T539& 20.0& 10.2 &1.75 10$^{11}$& 72& 27& 0.063$\pm $0.006&
22$\pm $3 \\
\hline T315& 10.0& 49.8 &2.30 10$^{11}$& 79& 10& 0.19$\pm $ 0.01&
6.0$\pm $0.2 \\
\hline NU211& 10.2& 32.8 & 3.10 10$^{11}$& 129& 13& 0.22$\pm
$0.01&
6.4$\pm $0.9 \\
\hline NU535& 6.8& 58.5 & 3.30 10$^{11}$& 138& 13& 0.29$\pm $0.02&
5.1$\pm $0.9 \\
\hline
\end{tabular}
\label{tab1}
\end{center}
\end{table*}

Figure 2 shows measurements of the photoluminescence (PL) and
photoluminescence excitation (PLE) spectra of the samples at 5 K
taken at the maximum value of $N_{S}$ permitted by the sample
design. In each case the PL peak corresponds to the transition
between the conduction (CB) and valence (VB) band extrema (see
inset) whereas the onset of PLE represents the interband
transition to the lowest unoccupied conduction band states at the
Fermi level. The energy difference between the PL peak and the PLE
onset was used to determine the electron concentration $N_{S}$
using electron and hole effective masses of 0.067 and 0.13
respectively. These $N_{S}$ values are more reliable than those
obtained from the Hall measurements and we have used them in our
analysis. The onset of the PLE shows a step or a peak
characteristic of the Mahan exciton which has been studied
previously in 2DEGs\cite{14}. When biased for lower $N_{S}$ the
spectra showed features characteristic of negatively charged
excitons~\cite{15} rather than 2DEGs. Consequently we were not
able to study effects of different concentrations in a true 2DEG
in a single sample. Nevertheless the variation of concentration
from sample to sample (Table 1) allows a direct test of
theoretical predictions (see section \ref{sec:3}). Table 1 also
contains measured values of the spin splitting at the Fermi
surface $\Omega (k_{F})$ and the electron momentum scattering time
$\tau _p^\ast $ for each sample, obtained from the study of the
spin dynamics of the samples in the quasi-collision-free low
temperature regime~\cite{5}. This data is used in our theoretical
analysis described in section~\ref{sec:3}. The value of $E_{e1}$
for each sample was calculated using PL and PLE spectra.

\begin{figure}[htbp]
\includegraphics[width=0.6\linewidth]{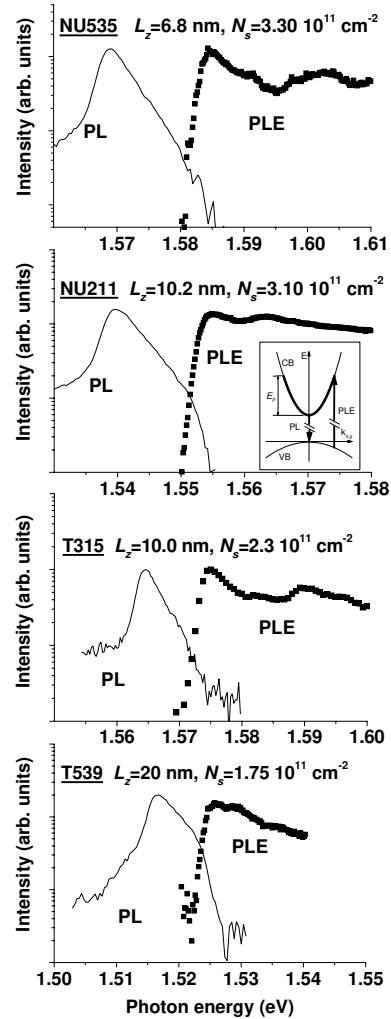}
\caption{Photoluminescence (PL) and photoluminescence excitation
(PLE) spectra of the different samples at 5 K. The measurements
were taken at the maximum value of $N_{S}$ permitted by the sample
design. The electron concentrations in the 2DEGs were determined
from the energy difference between PL peak and PLE onset which
involve interband transitions as indicated in the inset.
Time-resolved pump-probe measurements of spin dynamics were
carried out with the laser tuned to the PLE onset.} \label{fig2}
\end{figure}

The spin-dynamics of the 2DEGs were investigated using a
picosecond-resolution polarized pump-probe reflection
technique~\cite{16}. Wavelength-degenerate circularly polarized
pump and delayed linearly polarized probe pulses from a
mode-locked Ti-sapphire laser were focused at close to normal
incidence on the sample. Pump-induced changes of probe reflection
$\Delta R$ and of probe polarisation rotation $\Delta \theta $
were recorded simultaneously as functions of probe pulse delay
using balanced photodiode detectors and lock-in techniques. For
$\Delta R$, $10{\%}$ beam splitters allowed comparison of
intensities of the incident and reflected probe, and, for $\Delta
\theta $, a polarising beam splitter gave comparison of reflected
polarisation components at 45 degrees to the incident probe
polarisation. The pump beam intensity was typically 0.5 mW focused
to a 60 micron diameter spot giving an estimated photoexcited
spin-polarized electron density $5\times 10^{9}$ cm$^{-2}$, very
much less than the unpolarised electron concentration in the 2DEG
(Table 1); probe power density was $25{\%}$ of the pump.

At each temperature a wavelength excitation scan of the $\Delta R$
and {$\Delta \theta $} signals was recorded for a delay of 20 ps
and measurements of the time evolution were then made at the
wavelength of maximum signal. At low temperatures the maximum in
{$\Delta \theta $} coincided with the onset of the PLE spectrum in
each sample (see Fig. 2).

On the time scale of this experiment, phase-space filling by the
photoexcited electrons should dominate the pump-induced changes.
The $\sigma ^{+}$ circularly polarized pump photons will create an
excess population of $| S_{z}= 1/2 \rangle $ electrons at the
Fermi energy with isotropic distribution of in-plane wave vectors
and an equal population of $\vert J_{z}=-3/2\rangle$ holes in the
valence band. The phase-space-filling effect of the holes may be
neglected since they will rapidly relax into the lowest energy
states available at the top of the heavy hole valence band,
becoming depolarized at the same time. Furthermore, the majority
of optical transitions from these states will already be blocked
by the Fermi sea of electrons. The $\Delta \theta $ signal will
therefore be proportional to the pump-induced imbalance of
electron spin polarization along the growth axis $z$, $\langle
S_{z}\rangle$, and $\Delta R$ to the density of photoexcited
electrons.

\begin{figure}[htbp]
\includegraphics[width=0.8\linewidth]{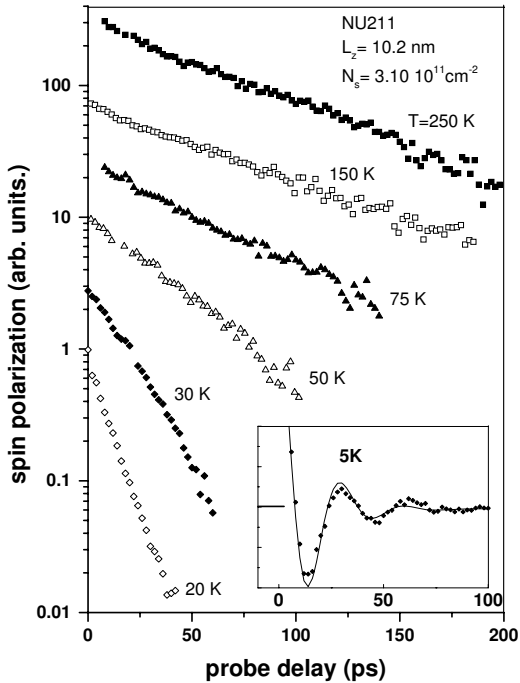}
\caption{Time evolution of the \textit{$\Delta \theta $ } signal
and hence the spin polarisation for NU211 sample for a range of
temperatures. At 20 K and above the evolution is exponential with
rapidly increasing decay time. At 5 K, oscillatory behaviour was
observed (inset) and was analysed by Monte-Carlo techniques
(solid)~\cite{5} to give directly $\Omega (k_{F})$ and $\tau
_p^\ast $.} \label{fig3}
\end{figure}

Figure 3 shows the time evolution of {$\Delta \theta $ }for sample
NU211 at several different temperatures; the traces are offset
vertically from one another for clarity. {$\Delta R$} was
essentially constant for this range of delays indicating
negligible decay of the photoexcited population. Thus the
behaviour of {$\Delta \theta $} indicates the pure spin-relaxation
of the electrons. It can be seen that the decay time increases
very rapidly as the temperature is raised, consistent with
decrease of $\tau _p^\ast $ in Eq. (\ref{eq1}) (the value of
$\langle\bm \Omega _{\bot }^{2}\rangle$ should not be strongly
temperature dependent, see section \ref{sec:3}). The inset shows
the behaviour of \textit{$\Delta \theta $} at 5K; the spin
evolution is oscillatory rather than exponential and analysis
using a Monte Carlo simulation technique\cite{4,5} gives the
frequency $\Omega (k_{F})$ and the scattering time $\tau _p^\ast
$~(Table 1).

\begin{figure}[htb]
\includegraphics[width=0.85\linewidth]{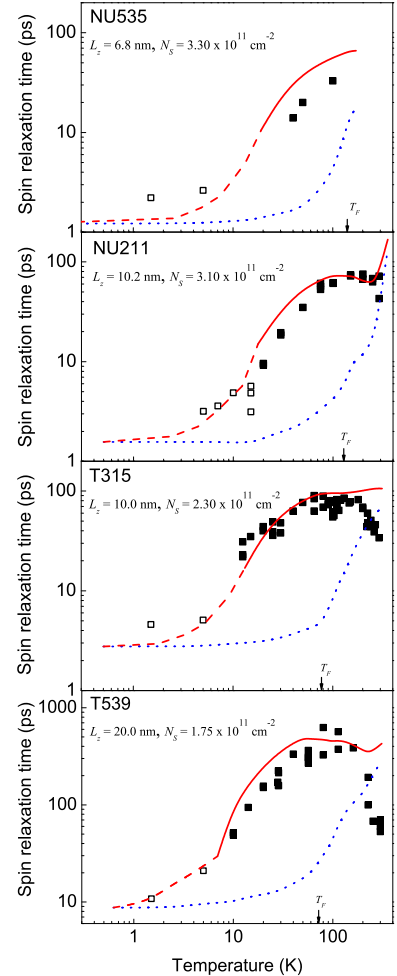}
\caption{Spin relaxation times vs. temperature. The points present
the experimental results; solid squares are directly measured
exponential decay times whereas open squares are values of
$[\Omega $(k$_{F})^{2}\tau _p^\ast]^{-1}$ for cases of oscillatory
spin evolution. Lines are the theoretical results calculated using
the experimental values of the ensemble momentum relaxation times
including (solid) and not including (dotted) electron-electron
collisions.} \label{fig4}
\end{figure}

Figure 4 shows the values of $\tau_s$ measured as a function of
temperature in the four samples. The solid symbols are from direct
measurements of exponential decay. The open symbols, at low
temperatures, do not represent exponential decay but are values of
$[\Omega(k_F)^2 \tau_p^*]^{-1}$ obtained from the observed
oscillatory spin evolution~\cite{5}. For each of the samples
$\tau_s$ increases rapidly with temperature and for the three
larger values of $L_{z}$ passes through a maximum at a temperature
which corresponds approximately to the Fermi temperature,
indicated by the arrows. For sample NU535 the same trend is
observed but there are insufficient points to identify a maximum
in the variation. There is also a clear trend towards lower $\tau
_{s}$ for smaller well width. The curves in Fig. 4 are the result
of our theoretical calculations described in section \ref{sec:3};
the solid and dashed curves represent calculated spin relaxation
times based on the DP mechanism and including both
electron-electron and ensemble momentum scattering whereas the
dotted curves are the same calculations but including only the
ensemble momentum scattering obtained from the measured mobility.

\section{ Discussion}\label{sec:3}

\subsection{Quantitative calculations}

Here we give an outline of the quantitative theory which we use to
interpret the experimental results; full details have already been
published in Refs. [\onlinecite{3}] and [\onlinecite{19}]. We use
the kinetic theory and describe spin-polarized electrons in the
framework of the spin-density matrix $\rho _{\bm k} =f_k +\bm
\sigma \bm s_{\bm k} $, where $f_k $ is the electron distribution
function and ${\bm s}_{\bm k} $ is the average spin of an electron
in the state with wave vector ${\bm k} $. Since the optical
excitation evidently makes no essential changes in the
distribution function $f_k $, it is enough to consider the time
evolution of the spin distribution function ${\bm s}_{\bm k} $
only. The spin distribution function satisfies the pseudovector
kinetic equation which, with allowance for electron-electron
interaction, reads~\cite{3}
\begin{equation}
\label{eq2}
 \frac{d {\bm s}_{\bm k}}{dt} +  {\bm s}_{\bm k} \times
\left( {\bm \Omega}_{\bm k} + {\bm \Omega}_{C\mbox{,}{\bm k}}
\right) + {\bm Q}_{\bm k} \{ {\bm s} \} = 0,
\end{equation}
where ${\bm Q}_{\bm k} \{ {\bm s}\}$ is the collision integral,
$\bm \Omega_{\bm k}$ and ${\bm \Omega}_{C\mbox{,}{\bm k}}$ are the
effective spin precession frequencies due respectively to the
spin-splitting and to the electron-electron exchange interaction.

In a quantum well there are three potential contributions to the
spin-splitting $\bm \Omega_{\bm k}$ (Refs.~[\onlinecite{7,17}])
originating from inversion asymmetry of the crystal structure
(Dresselhaus or BIA), from asymmetry of the quantum well structure
and applied odd parity perturbations such as electric field
(Rashba or SIA) and from asymmetry of atomic arrangement at
interfaces (NIA). The NIA term may be ignored for GaAs/AlGaAs
structures~\cite{7}. The BIA and SIA contributions each have a
leading term linear in $k$ giving total e1 subband splitting
\begin{equation}
\label{eq3} \frac{\hbar }{2}\left( {\bm \sigma }\cdot {\bm \Omega
}_{\bm k}  \right)=\beta _1 (\sigma _y k_y -\sigma _x k_x )+\beta
_2 (\sigma _x k_y -\sigma _y k_x )\:,
\end{equation}
where $\beta _1$, $\beta _2 $ are the constants describing BIA and
SIA contributions respectively and $x$ and $y$ are the
crystallographic axes [100] and [010]. The distinctive feature of
the BIA term is the factor $\langle k_{z}^{2}\rangle$ in
$\beta_1$, where $k_{z}$ is the wavevector component along the
growth axis~\cite{12}. We note that it arises as a quantum
mechanical average of the cubic-in-$\bm k$ splitting in bulk
zince-blende semiconducturs and $\langle k_{z}^{2}\rangle$ is
proportional to the electron confinement energy, $E_{e1}$.

 The electron-electron exchange
splitting is given by
\begin{equation}\label{eq4}
{\bm \Omega}_{C\mbox{,}{\bm k}} = \frac{2}{\hbar} \sum_{\bm k'}
V_{\bm k' - \bm k} \bm s_{\bm k'}\:,
\end{equation}
where $V_{\bm q}$ is the Fourier transform of the screened 2D
Coulomb potential. The full expression for the electron-electron
scattering contribution to the collision integral is given in
Refs.~[3,19] (see also Ref.~[20]) and is not presented here. The
momentum relaxation processes governing the electron mobility are
described in the framework of the temperature-dependent scattering
time $\tau _p (T)$.

For the samples under study the electron polarisation $P_s$ is of
order of $1\%$ and the $\bm \Omega_{C,\bm k}$ (Hartree-Fock) term
in Eq.~(\ref{eq2}) can be ignored~\cite{3}.


Assuming $\Omega _k \tau _p^\ast \ll 1$, which is valid for the
temperatures exceeding 10 K to 20 K depending on the sample,
Eq.~(\ref{eq2}) can be solved by iteration in this small parameter
leading to exponential decay of the total electron spin: $\bm
S(t)=\bm S_0 \exp (-t/\tau _s )\parallel z$. While solving
Eq.~\eqref{eq2} the spin distribution function $\bm s_{\bm k}$ is
represented as a sum ${\bm s}^0_{\bm k} + \delta {\bm s}_{\bm k}$
where $\bm s^0_{\bm k}$ is the quasi-equilibrium function, and the
correction $\delta {\bm s}_{\bm k} \propto \Omega_{\bm
k}\tau_p^\ast$ arises due to the spin precession. This correction
is determined from the linearized equation with the
result~\cite{3}
\begin{equation} \label{eq5}
\delta {\bm s}_{\bm k} = F_k\: {\bm S}(t) \times {\bm \Omega}_{\bm
k} \:,
\end{equation}
where $F_k$ is a function independent of the wave-vector
direction. Finally, one can arrive to the following expression for
the $zz$ component of the tensor of spin relaxation rates
\begin{equation}
\label{eq6} \frac{1}{{\tau _s }} = \sum\limits_{\bm k}  \left|
{{\bm{\Omega }}_{\bm{k}}^{} } \right|^2 F_k  = \frac{4}{{\hbar ^2
}}(\beta _1^2 + \beta _2^2 )\sum\limits_{\bm k}  \,k^2 F_k \:.
\end{equation}

In Fig.~4 we show theoretical curves for $\tau _s$ calculated
using Eq. \eqref{eq6}; the solid and dashed curves were calculated
taking into account both electron-electron collisions and the
ensemble scattering processes using the measurements of $\tau_p
(T)$ (Fig. 1); the dotted curves were calculated including only
the ensemble scattering. In the high temperature regime (solid
portion of curve) where $\Omega_k \tau _p^\ast \lesssim 1$ the
theory predicts exponential spin relaxation and values of $\tau_s$
can be compared directly with the experimental decay times (solid
symbols).  In the low temperature regime where $\Omega_k \tau
_p^\ast \gtrsim 1$ the spin dynamics become oscillatory and
quantities calculated using Eq. \eqref{eq6} (dashed portion of
curve) do not represent exponential decay times. They can however
be directly compared with the values of  $[\Omega (k_{F})^{2}\tau
_p^\ast]^{-1}$ (open symbols) obtained from the observed
oscillatory spin evolution.

The calculations have no adjustable parameters. The parameters
used are all determined independently and are given in Table 1;
they are well width ($L_{z})$, barrier height (determined by Al
concentration), scattering time $\tau _p (T)$ (from mobility),
electron concentration ($N_{S})$ and the spin splitting constant
$\sqrt {\beta _1^2 +\beta _2^2 }$ [obtained from the measured
values of $\Omega $(k$_{F})$]. From the PL peak positions in
Fig.~2 it is clear that the true well widths, particularly for
NU211 and T315 are somewhat different than the nominal ones. The
well-width enters directly only in the calculation of the
electron-electron scattering; although we used the nominal values
of well width the errors involved will be small because the
electron-electron scattering times are not strongly dependent on
well width~\cite{2}. The well width also has a stronger indirect
effect via the value of $\Omega(k_F)$ which is measured for each
sample.

\subsection{Qualitative interpretation of the spin relaxation}

Before considering in detail the comparison of the calculations
with experiment we give qualitative arguments which provide a
physical understanding of the temperature, concentration and
well-width dependence of the spin relaxation time shown in Fig. 4.
Returning to Eq. (\ref{eq1}), we see that the spin relaxation time
is inversely proportional to the product of the average squared
spin precession frequency and the scattering rate.

Let us first consider $\left\langle {\bm \Omega _\bot ^2 }
\right\rangle $. Both SIA and BIA terms are linear in the in-plane
wavevector. This means that at relatively low temperatures where
electron confinement energy exceeds the thermal energy, to a first
approximation $\bm \Omega_{k }$ is linear in the in-plane electron
wavevector and $\left\langle {\bm \Omega _\bot ^2 } \right\rangle$
is proportional to the average in-plane kinetic energy. For a
degenerate electron gas the latter is independent of $T$ and
proportional to Fermi temperature $T_{F } \sim N_S$ (we remind
that $T_F = E_F /k_B$). For BIA spin splitting at temperatures
less than $E_{e1}$/$k_{B}$, $\left\langle {\bm \Omega _\bot ^2 }
\right\rangle $ should be approximately proportional to
$E_{e1}^{2}E_{F }$. This tendency is consistent with the data
given in Table 1, see also Ref.~\onlinecite{5}. For a
non-degenerate 2DEG $\left\langle {\bm \Omega _\bot ^2 }
\right\rangle $ is linear in $T $ and independent of $ N_{S}$.
Note, that in the temperature range of our experiments
$\left\langle {\bm \Omega _\bot ^2 } \right\rangle $ changes by no
more than a factor of 3.

Now consider the scattering rate $({\tau _p^\ast})^{-1}$. If we
ignore electron-electron scattering, $({\tau _p^\ast})^{-1}$ will
follow the inverse of the mobility (see Fig. 1), constant at low
temperatures and increasing roughly as $T^{2}$ at high
temperatures where phonon scattering takes over. Combining this
with the variation of $\langle \bm \Omega _{\bot
}^{2}\rangle^{-1}$ gives a contribution to the temperature
dependence of $\tau_{s}$ which is constant at low temperature and
roughly proportional to $T$ at high temperatures. This is
indicated by the dashed curve in Fig.~5, which follows
qualitatively the dotted curves in Fig.~4.

Our next step is to include the electron-electron scattering.
Physically, the collisions between electrons change randomly the
orientation of the wavevector of a given electron and, if the
spins of colliding electrons are different, leads to randomization
of the precession frequencies $\bm \Omega_{\bm k}$ exactly as for
other scattering processes. For a degenerate electron gas the
electron-electron scattering rate of an electron near the Fermi
energy is governed by the Pauli-exclusion principle. Phase-space
arguments~\cite{18} demonstrate that for a pair of electrons the
number of free final states is proportional to the squared ratio
of temperature and Fermi energy. According to
Ref.~[\onlinecite{3}] the effective scattering rate which is
relevant to the D'yakonov-Perel' spin relaxation mechanism has,
for a strictly 2D system, the form
\begin{equation} \label{eq7}
\frac{1}{\tau _{ee} }\approx 3.4\frac{E_F }{\hbar }\mathop {\left(
{\frac{k_B T}{E_F }} \right)}\nolimits^2 \:,\mbox{ }T\ll T_F .
\end{equation}
The Fermi level $E_{F}$ is proportional $N_{S}$ giving
\begin{equation} \label{oldEq3}
\tau_{ee}^{-1}\sim T^{2}N_{S}^{-1} \quad T \ll T_{F}.
\end{equation}
This counter-intuitive concentration dependence arises because,
for a fully degenerate electron gas, i.e. at $T$=0 K, the
electron-electron scattering rate vanishes due to Pauli exclusion
principle. At a finite temperature the rate becomes finite but now
any change which moves the system back towards full degeneracy,
such as increase of electron concentration, will produce a
reduction in the scattering rate. For a nondegenerate electron gas
the electron-electron scattering rate is determined by the
wavevector dependence of the Coulomb matrix element and the
density of electrons. Thus~\cite{3},
\begin{equation} \label{eq8}
\tau _{ee}^{-1} \approx 35.7\frac{e^4 N_S }{\hbar
\mbox{{\ae}}^2k_B T}\:\sim N_S T^{-1},\mbox{ }T\gg T_F.
\end{equation}
Combining expressions \eqref{oldEq3} and (\ref{eq8}) with
$\langle\bm \Omega _{\bot }^{2}\rangle^{-1}$ gives
\begin{equation}\label{oldEq5}
\tau_{s} \sim \left\{
\begin{array}{cc}
T^{2}N_{S}^{-2},& T \ll T_{F},\\
T^{-2}N_{S} ,& T\gg T_{F }.
\end{array}
\right.
\end{equation}
These contributions for one value of $N_{S}$ are shown as the
dotted curve in Fig. 5 (the region near $T=T_{F}$ being no more
than a guide to the eye) and the combined effect of the different
scattering processes is given by the solid curve.

\begin{figure}[htbp]
\includegraphics[width=0.8\linewidth]{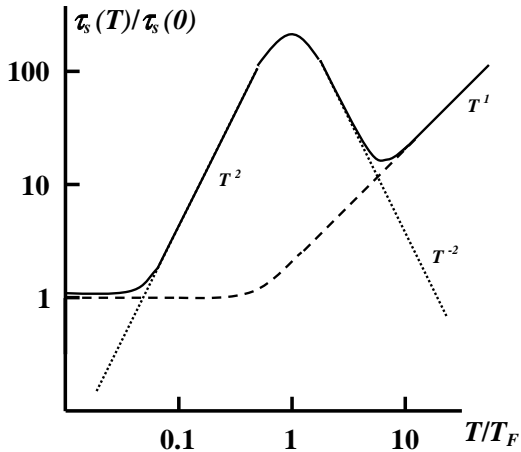}
\caption{Qualitative picture of the temperature dependence of the
spin relaxation time. Dashed curve presents the contribution of
the ensemble momentum scattering processes. Dotted curve is the
prediction for electron-electron collisions only. The solid curve
is the combined total spin relaxation time.} \label{fig5}
\end{figure}

Clearly these considerations give no more than a very crude
qualitative picture but nonetheless contain the essential physical
explanation of the observed rapid increase with temperature of the
spin relaxation time as well as the observed maxima for the
different samples. On the assumption that the BIA term is dominant
we expect that $\left\langle {\bm \Omega _\bot ^2 } \right\rangle
\sim E_{e1}^{2 }$ and therefore, other factors remaining constant,
that the spin relaxation time at a given temperature will scale as
$E_{e1}^{-2}$ or approximately as $L_{z}^{4}$. This is a somewhat
stronger dependence than we observe experimentally as can be seen
from Fig. 4. The actual behaviour must be understood by inclusion
also of the differences of $\tau_{p}^\ast$ and in the relative
importance of the SIA and BIA terms for different samples.

\subsection{Comparison of quantitative calculations with experiment}
It is clear from Fig.~4 and the arguments leading to Fig.~5 that a
proper description of the temperature dependence of the spin
relaxation requires inclusion of electron-electron scattering;
calculation neglecting electron-electron collisions does not fit
the experimental points but agreement when including
electron-electron scattering is excellent. In particular the
calculation reproduces very well the observed rapid increase of
spin relaxation time with temperature between 10K and 100K and
also the observed maxima which occur close to the transition
temperature $T_{F}$ between degenerate and nondegenerate regimes
of the 2DEG.

The only significant disagreement between theory and experiment is
above about 200~K in samples T315 and T539; here the trend of the
theory is upwards with temperature due to the fact that, in this
range, the ensemble momentum scattering rate is dominating over
the electron-electron scattering; at the same time the
experimental points are falling with temperature. This could
indicate the presence in the samples of some spin-relaxation
process other than the DP mechanism or, if the explanation stays
within the DP mechanism (Eq (\ref{eq1})) either that total
electron momentum scattering is weaker than we have assumed or
that the precession term $\left\langle {\bm \Omega _\bot ^2 }
\right\rangle$ is stronger. Below we briefly discuss these
possible reasons for the discrepancy.

First, we exclude spin relaxation mechanisms other than DP. The
Bir-Aronov-Pikus (BAP) mechanism\cite{10,6,7} is irrelevant as
electrons are the major carriers in the samples under study. We
believe also that paramagnetic impurities~\cite{23} play no role
as there is no evidence of their presence and furthermore such a
scattering cannot result in strong temperature dependence of the
spin relaxation time as we observe here. The Elliott-Yafet (EY)
mechanism~\cite{8,9,6,7,24,25}, where the spin relaxation occurs
due to spin-flip scattering is unimportant at high temperatures
even in bulk GaAs and is further suppressed in quantum wells. For
an estimate of the spin relaxation time one may use $\tau _s
(T)\sim 10^5\tau _p (T)$ (see Ref.~[\onlinecite{26}]), which
gives, for our lowest value of $\tau _p \sim 10^{-13}$~s (Fig.~1),
$\tau _s \sim 10^{-8}$~s, at least two orders of magnitude larger
than the experimental values.

Thus, we come to the conclusion that the spin relaxation in the
temperature range $200$~K to $300$~K is governed by the DP
mechanism. One way in which we could have underestimated
$\left\langle {\bm \Omega _\bot ^2 } \right\rangle $ is if our
assumption of temperature-independence of $\beta _1^2 +\beta _2^2
$ in Eq.~(\ref{eq6}) were to break down. In principle this could
occur if, in the high-temperature regime, built-in electric fields
vary in the sample causing the coefficient of the SIA term
$\beta_2$ to play an increasing role. However, estimates of the
electric field required to explain the effect ($\sim 100$
kVcm$^{-1}$) make this unlikely.

Finally we examine the possibility that the discrepancy can be
explained by an overestimation of the electron scattering rate
used in the calculations. In the worst case, T539 at 300 K (Fig.
4), we see that not only is a tenfold reduction in the overall
scattering rate required but also, since the dotted curve lies
above the experimental points, a reduction of the ensemble
scattering rate by at least a factor 3 is required, implying that
our Hall measurements have underestimated the mobility by that
factor at 300~K. The remaining discrepancy would then disappear if
the well were completely depleted of electrons by 300~K rather
than the electron concentration being constant, as we assume in
calculating the solid curves. We cannot rule out such an error in
the mobility since we are unable to assess the effects of
transport in other parts of the sample than the 2DEG and,
furthermore, as mentioned in section \ref{sec:2}, the Hall
measurements do indicate some reduction of concentration at higher
temperatures. Therefore it is most likely that the discrepancies
at high temperature in Fig. 4 are due to errors in the mobility
and concentration measurements and do not imply a fundamental lack
of physical understanding of the system.

\section{Conclusion}

We have made a comprehensive experimental and theoretical study of
the electron spin relaxation in 2DEGs in quantum wells with
different widths, carrier concentrations and carrier mobilities.
The main observations are of rapid increase of the spin-relaxation
time between 10~K and 100~K to a maximum at a temperature close to
the Fermi temperature of the 2DEG. The main spin relaxation
mechanism, namely, the Dyakonov-Perel' mechanism is identified for
all the samples in the whole temperature range under study. We
have made Hall measurements of electron mobility to determine the
momentum scattering rate for the electron ensemble and
measurements of the low temperature spin beats~\cite{5} allowed us
to directly determine the spin precession rate at the Fermi level.
It turns out that the elastic momentum scattering processes which
govern the electron mobility \textit{do not} play an important
role in the spin relaxation for the temperature range 10 K -- 150
K. The only possible candidate is thus \textit{electron-electron
collisions} which conserve the total momentum of the pair but
contribute to the randomization of the spin precession
frequencies. The experimental results and our theoretical
calculations (made without fitting parameters) are in good
agreement with each other. We conclude that the spin dynamics in
high-mobility $n$-type quantum wells is determined by the
Dyakonov-Perel' (or precessional) spin relaxation mechanism,
governed by electron-electron collisions.

\acknowledgements

The work was supported finacially by RFBR, ``Dynasty'' foundation -- ICFPM,
the Engineering and Physical Sciences Research Council (EPSRC) and the Royal
Society.

\end{document}